\documentclass{article}

\usepackage{arxiv}
\usepackage[T1]{fontenc}
\usepackage{amsmath,amssymb}
\usepackage{newtxtext,newtxmath}
\usepackage{microtype}
\usepackage{cite}
\usepackage{booktabs}
\usepackage{array}
\usepackage{tabularx}
\usepackage{graphicx}
\usepackage{tikz}
\usepackage{xcolor}
\usepackage{xurl}
\usepackage[hidelinks]{hyperref}
\usetikzlibrary{arrows.meta,positioning,fit,calc}

\definecolor{agentblue}{RGB}{36,91,141}
\definecolor{agentgold}{RGB}{201,142,22}
\definecolor{softblue}{RGB}{232,241,248}
\definecolor{softgold}{RGB}{252,244,218}
\newcolumntype{L}[1]{>{\raggedright\arraybackslash}p{#1}}
\newcolumntype{Y}{>{\raggedright\arraybackslash}X}

\title{Governed AI-Agent Coordination for Dementia Care:\
Architecture, Safety Contracts, and Evidence-Derived Workflow Verification}

\author{
  Francesca Medda \quad Hui Gong\\
  UCL Institute of Finance and Technology\\
  University College London\\
  London, United Kingdom
}

\date{October 2026}
\renewcommand{\headeright}{Preprint}
\renewcommand{\undertitle}{Preprint}
\renewcommand{\shorttitle}{Governed AI-Agent Coordination for Dementia Care}

\hypersetup{
  pdftitle={Governed AI-Agent Coordination for Dementia Care: Architecture, Safety Contracts, and Evidence-Derived Workflow Verification},
  pdfauthor={Francesca Medda and Hui Gong},
  pdfkeywords={AI agents, dementia care, agentic engineering, persistent memory, human-AI collaboration, AI governance}
}

\begin{document}
\makeatletter
\immediate\write\@auxout{\string\citation{IEEErefControl}}
\makeatother
\maketitle

\begin{abstract}
Dementia care at home increasingly involves connected sensors, medication devices, electronic records, and assistive technologies. Professionals and family carers must still interpret these observations, reconstruct context, coordinate responses, and verify resolution. Sustaining safe care at home while ensuring timely clinical escalation requires continuity between these tasks. Agentic engineering creates a systems opportunity by combining model interpretation with persistent external memory, planning, authorised tool use, and outcome feedback. This paper presents Governed Closed-loop Agent Coordination (GCAC), an architecture for bounded agent participation in community dementia-care workflows. Evidence on coordination failures and policy obligations is translated into traceable system requirements. GCAC separates observation, four governed memory classes, planning, deterministic policy enforcement, execution, and outcome monitoring through a typed event--memory--decision--action--outcome contract. A worked medication-device example explains how these roles maintain context, allocate responsibility, and distinguish an unrecorded dose from a clinically established omission. A reference harness evaluates 18 evidence-derived traces spanning care gaps, service failure, consent change, stale state, duplicate events, untrusted text, and acute-change routing. GCAC satisfies all 18 contract oracles with zero policy-violating tool calls; event-threshold and stateless-planner controls satisfy 2/18 and 1/18, respectively. Six ablations identify the contribution of memory, policy, and versioning. The results establish architectural conformance. Effects on hospital use, quality of life, staff workload, unpaid care, and net costs require prospective evaluation. GCAC provides a testable basis for accountable coordination that preserves human authority over consequential care decisions.
\end{abstract}


\section{Introduction}
Dementia creates changing cognitive, functional, behavioural, and social-support needs that cross the home, primary care, specialist services, hospitals, pharmacies, and community organisations. The World Health Organization identifies coordinated diagnosis, treatment, care, support, and information systems as a public-health priority \cite{WHO2021Dementia}. In the United Kingdom, approximately 982,000 people were estimated to be living with dementia in 2024, with annual economic and social costs of about \pounds42 billion; unpaid care accounted for approximately half of the estimated total \cite{AlzSoc2024}. Coordination therefore consumes both professional capacity and family time: noticing change, locating missing information, reconciling different accounts, contacting the appropriate person, following up a request, and recording what happened.

Digital technologies have long supported dementia care through sensing, monitoring, communication, and assistance. Smart-home sensors, wearables, mobile applications, speech systems, and socially assistive robots can monitor activity, prompt routines, facilitate communication, or provide companionship. Reviews nevertheless identify fragmented interfaces, installation and configuration burden, limited pathway integration, uneven accessibility, and heterogeneous evidence \cite{Cornelius2025,Xie2020,Fan2025,Nam2025}. A door sensor reports that a door opened; a medication device may report that a dose was not recorded; a record interface exposes a prescription list. Context is needed to establish whether intervention is necessary, which response is proportionate, and who should act. A missing device record alone does not establish a medication omission.

The care setting gives this coordination problem practical importance. UCL's 2026 report \emph{Breaking the cycle of hospital admissions in people with dementia} describes adverse outcomes associated with hospitalisation, including delirium, cognitive and functional decline, and increased mortality, alongside communication gaps and delayed discharge \cite{UCL2026Hospital}. Supporting safe care at home therefore requires continuity between everyday observations, current care arrangements, and appropriate professional responses. Preventing avoidable crises is a service objective that must coexist with timely hospital treatment when needed; an admission-avoidance target must never obstruct a necessary escalation. Whether technological coordination improves either outcome requires direct evaluation.

This gap is measurable rather than merely conceptual. In a study of people living with dementia who had fragmented ambulatory care, Kern \emph{et al.} found that 57\% of respondents reported at least one coordination problem and 18\% reported an adverse event attributed to poor coordination \cite{Kern2024}. Reported gaps included a primary doctor not being current on specialist care, failure to follow up test results, incomplete medication review, and unavailable records. Reported adverse events included repeated tests and drug--drug interactions involving multiple prescribers. Patients and caregivers are often the first to recognise these failures, but their observations are not consistently converted into tracked service tasks.

The central design problem is to sustain an authorised workflow across time and organisational boundaries. Earlier Internet-of-Things (IoT) integration and rule engines could exchange messages and automate predefined event--action pairs. A multimodal large language model (LLM) can interpret richer input and draft a recommendation, but a model invocation alone does not establish durable operational memory, sustained objectives, an executable workflow, or explicit accountability. Without an external state-and-control mechanism, people must still arrange approval, implement a response, and determine whether the concern has been resolved.

Agentic engineering changes this system boundary. A runtime can retrieve and update external memory, decompose a goal, invoke tools, inspect their responses, and continue until an episode is completed, transferred, or safely stopped. ReAct demonstrated the value of interleaving reasoning and action \cite{Yao2023}; generative-agent and Reflexion architectures illustrated how external episodic memory, retrieval, and feedback can support behaviour across time \cite{Park2023,Shinn2023}; and multi-agent runtimes now compose models, people, and tools \cite{Wu2024AutoGen}. These capabilities are unusually relevant to dementia care because the meaning of a current observation often depends on personal routine, earlier events, current capacity and consent, an evolving care plan, and unresolved actions. Memory in this setting is not an anthropomorphic substitute for a person's memory. It is an accountable representation of care continuity.

The opportunity lies in bounded agent participation in governance and coordination: maintaining the latest authorised state; citing the evidence used; proposing a proportionate next step; obtaining human approval where required; invoking only an allow-listed tool; and preserving an outcome or unresolved obligation. Recent reviews describe goal-directed tool use and workflow orchestration as emerging healthcare-agent capabilities, while also emphasising limited deployment evidence, inconsistent evaluation, and the need for stronger safety and governance \cite{Zhao2026,Collaco2026,Ferber2026}. This paper addresses the systems layer between model capability and a care pathway.

Governed coordination can be understood as an investment in the organisation of care. Its proposed value comes from preserving context, reducing duplicated activity, making responsibility explicit, and keeping unanswered requests visible. Those mechanisms may reduce repeated coordination by professionals and families, but economic value depends on their effects relative to implementation, maintenance, and oversight costs. This motivates an architecture whose responsibilities and outcomes can be inspected before any claims about health or financial benefit are tested.

We ask three research questions. \emph{RQ1:} can a governed agent runtime complete an event-to-outcome care-coordination contract across evidence-derived workflow traces? \emph{RQ2:} which memory and governance components are necessary for those contracts to hold? \emph{RQ3:} what failure signatures appear when those components are absent? We make four contributions:
\begin{enumerate}
    \item an evidence-to-requirement derivation linking documented dementia-care coordination gaps, person-centred care obligations, and agent-system controls;
    \item Governed Closed-loop Agent Coordination (GCAC), a runtime architecture that separates generative planning from deterministic authority and represents closure as a first-class outcome;
    \item a typed event--memory--decision--action--outcome contract with four governed memory classes, versioned state transitions, action authority levels, and explicit failure containment; and
    \item an executable reference harness with 18 requirement-derived traces, two controlled baselines, six component ablations, quantitative results, and error analysis.
\end{enumerate}

The evaluation is deliberately architectural. It uses no patient data, does not estimate diagnostic accuracy, and does not claim improved health outcomes. Its purpose is to falsify coordination and governance claims before any clinical or social-care deployment.

\section{Evidence, Related Work, and Requirements}

\subsection{Dementia Technology and the Coordination Gap}
Digital dementia-care research includes monitoring, prompting, cognitive support, caregiver support, remote communication, and robot-assisted interventions \cite{Cornelius2025,Xie2020,Fan2025,Nam2025}. Much of this work evaluates a device, interface, or intervention in isolation. That evidence is important but leaves a systems question unanswered: how should observations and capabilities from heterogeneous devices, people, and services become one accountable episode of care?

The coordination findings of Kern \emph{et al.} provide concrete failure modes \cite{Kern2024}. Among 88 respondents exposed to fragmented ambulatory care, 35\% reported receiving insufficient help from the personal doctor's office to manage care across providers; 25\% believed doctors did not communicate; 20\% reported that the personal doctor was not consistently up to date on specialist care; 18\% reported insufficient test-result follow-up; 14\% reported incomplete discussion of prescription medicines; and 6\% reported missing records at a scheduled appointment. These figures come from one US accountable-care setting and are not prevalence estimates for the United Kingdom. Their value here is requirements evidence: each identifies information that must persist, a task that must be assigned, or an outcome that must be verified.

The study also supports a patient- and caregiver-reporting channel. A care platform that ingests only sensors and clinical records can still miss the person who knows that two providers have not communicated, that a result never arrived, or that a promised call did not occur. GCAC therefore treats person- and caregiver-reported concerns as provenance-bearing observations rather than informal free text outside the workflow. Such input can open a review task; it cannot itself alter permissions or prove a clinical condition.

The connection between coordination and clinical benefit requires its own evidence. Howard \emph{et al.}'s umbrella review found high-certainty evidence that case management did not reduce hospitalisation; evidence for virtual case management also failed to establish a consistent reduction. Clinical pharmacists within multidisciplinary teams and advance care planning had more promising evidence at different certainty levels \cite{Howard2026Hospital}. These findings motivate support for specific professional workflows and documented care preferences. They do not establish that adding agents to coordination will reduce hospital use.

\subsection{From Agents as Models to Agents as Governed Systems}
Software agents predate LLMs and have long been described as autonomous components situated in an environment and capable of flexible action \cite{Jennings2000}. What has changed is the practical combination of foundation-model interpretation, multimodal interfaces, external memory, tool APIs, and orchestration frameworks. An LLM can now translate heterogeneous observations into structured candidate actions; an agent runtime can maintain goals and state; and standardised service interfaces can expose communication, scheduling, record retrieval, and assistive-device functions.

These capabilities do not automatically create trustworthy agency. A longer model context is not durable memory: it lacks independent provenance, retention, concurrency control, and purpose limitation. A multi-agent conversation is not governance: agents can agree on an action that none is authorised to execute. A tool call is not workflow completion: delivery can fail, the recipient can be unavailable, and an acknowledgement can be absent. GCAC therefore treats the foundation model as a replaceable planner within a larger state machine. The contribution lies in the care-specific contracts around the model, not in claiming a new foundation model.

Healthcare-agent reviews similarly distinguish potential from evidence. Applications are expanding, but evaluation often focuses on task accuracy rather than longitudinal workflow integrity, tool authority, outcome closure, or the distribution of responsibility between people and machines \cite{Zhao2026,Collaco2026,Strong2026}. GCAC makes those properties explicit and testable.

\subsection{Policy and Governance Context}
NICE guidance requires person-centred decisions that reflect needs, preferences, and values and appropriately involve families and carers \cite{NICE2018}. In England and Wales, capacity is decision-specific and may fluctuate; a system must not treat a dementia diagnosis as a permanent, global removal of agency \cite{MentalCapacityAct2005}. WHO guidance frames autonomy, safety, transparency, accountability, equity, and sustainability as system requirements for AI in health \cite{WHO2021AI}.

GCAC translates these principles into three linked control domains. \emph{Data governance} addresses source reliability, provenance, necessary processing, lawful basis, retention, and relevant consent. \emph{Model governance} tracks model, prompt, retrieval, and calibration changes. \emph{AI-system governance} assigns responsibility for proposals, approvals, execution, escalation, and closure. These domains meet at the action boundary: a well-formed recommendation still requires the authority appropriate to its purpose, recipient, and consequences.

The UK's National Commission into the Regulation of AI in Healthcare now recommends proportionate lifecycle oversight, meaningful human involvement, system-wide responsibility, transparency, staged deployment, and continuing real-world monitoring \cite{MHRA2026Recommendations}. These principles matter even when a coordination tool is not itself used for diagnosis. The relevant unit of assurance is the combined system of models, software, data, tools, organisations, and users. NIST's AI Risk Management Framework and generative-AI profile similarly emphasise governed, mapped, measured, and managed risk across the lifecycle \cite{NIST2023,NIST2024GenAI}.

The emerging National Care Service agenda adds an operational policy direction. Its stated principles include seamless movement between hospitals, home care, and community services, prevention, support for independent living, and cooperation across local social, primary, and community care \cite{UKGov2026Care}. GCAC is not a proposal for a central national decision-making agent. It is a reference pattern for preserving authorised tasks and outcomes as responsibility moves across those boundaries.

\subsection{Evidence-to-Requirement Derivation}
Table~\ref{tab:requirements} translates observed problems and governance obligations into architecture requirements and testable properties. Every trace in Section~\ref{sec:method} maps to at least one documented care gap, policy obligation, interface failure, or security hazard. The hospital-care evidence motivates the service setting; the finite trace suite tests coordination contracts within that setting.

\begin{table}[t]
\caption{Evidence-to-Requirement Traceability}
\label{tab:requirements}
\centering
\footnotesize
\begin{tabularx}{\textwidth}{L{0.22\textwidth}L{0.25\textwidth}Y Y}
\toprule
Evidence or obligation & Failure if untreated & GCAC requirement & Verifiable property \\
\midrule
Specialist updates and records unavailable \cite{Kern2024} & Decisions use incomplete state; requests disappear between organisations & Versioned shared care state and unresolved-task tracking & Missing update remains assigned until supplied, transferred, or explicitly closed \\
Test results not followed up \cite{Kern2024} & Repeat tests, delay, and ambiguity about responsibility & Outcome monitor with deadline and acknowledgement & Delivery is not equated with result receipt; absent outcome remains open \\
Medication conflict across prescribers \cite{Kern2024} & A model or device acts beyond authority & Reconciliation proposal plus named professional approval & Medication alteration has no autonomous tool binding \\
Patient/carer detects a coordination gap \cite{Kern2024} & Valuable evidence remains outside the record and workflow & Provenance-bearing reporting channel & Concern opens a reviewable task but cannot alter policy \\
Person-centred choice, consent, and fluctuating capacity \cite{NICE2018,MentalCapacityAct2005} & Static or generic action conflicts with current wishes or lawful authority & Time-, purpose-, action-, and recipient-specific consent state & Revoked purpose is denied; consequential action requires current authority \\
Human oversight and system-wide assurance \cite{WHO2021AI,MHRA2026Recommendations} & Generative recommendation becomes de facto permission & Deterministic policy gate and role-specific action levels & A3 requires signed approval; A4 is non-executable \\
Distributed messaging and changing records & Duplicate action or overwrite of newer care state & Idempotency keys and optimistic concurrency & Replayed event executes once; stale state is rejected \\
Untrusted records and free text \cite{NIST2024GenAI} & Embedded instruction changes tools, contacts, or permissions & Data/instruction separation and least-privilege tools & Free text cannot modify semantic policy state \\
Service or channel failure & Sent message is mistaken for completed care & Explicit fallback owner and terminal-state contract & Failure becomes approved fallback or visible unresolved obligation \\
\bottomrule
\end{tabularx}
\end{table}

\section{Governed Closed-loop Agent Coordination}

\subsection{System Boundary and Design Questions}
GCAC is designed around three questions that a connected-device stack does not answer. First, \emph{what information must persist?} A current event must be interpreted against routine, consent, earlier episodes, and unfinished work. Second, \emph{who is allowed to decide and act?} Interpretation, recommendation, permission, execution, and clinical judgement are different functions. Third, \emph{how does the system know that an episode is closed?} A sent message is not an acknowledged contact, and an acknowledged contact is not necessarily a resolved task.

\begin{figure}[t]
\centering
\resizebox{0.96\textwidth}{!}{%
\begin{tikzpicture}[
box/.style={draw=agentblue, rounded corners=1.5pt, fill=softblue, align=center, minimum height=8mm, text width=27mm, font=\footnotesize},
gold/.style={draw=agentgold, rounded corners=1.5pt, fill=softgold, align=center, minimum height=8mm, text width=29mm, font=\footnotesize},
label/.style={font=\footnotesize\bfseries, anchor=east},
arr/.style={-{Latex[length=1.7mm]}, thick, agentblue}, node distance=4mm]
\node[label] (l1) at (0,0) {Connected devices};
\node[box, right=of l1] (d1) {Sensors and devices};
\node[box, right=of d1] (b1) {Broker and fixed rules};
\node[box, right=of b1] (a1) {Alerts and dashboard};
\node[gold, right=of a1] (h1) {Human reconstructs, decides, acts, follows up};
\draw[arr] (d1)--(b1); \draw[arr] (b1)--(a1); \draw[arr] (a1)--(h1);

\node[label, below=7mm of l1] (l2) {Model-assisted};
\node[box, right=of l2] (d2) {Multimodal input};
\node[box, right=of d2] (c2) {Supplied context};
\node[box, right=of c2] (llm2) {Foundation model};
\node[gold, right=of llm2] (h2) {Recommendation; human executes and verifies};
\draw[arr] (d2)--(c2); \draw[arr] (c2)--(llm2); \draw[arr] (llm2)--(h2);

\node[label, below=7mm of l2] (l3) {Governed agentic};
\node[box, right=of l3] (d3) {Devices, records, people};
\node[box, right=of d3] (m3) {Versioned care memory};
\node[box, right=of m3] (ag3) {Bounded agent roles};
\node[gold, right=of ag3] (g3) {Policy gate and named human authority};
\node[box, right=of g3] (t3) {Tools and verified outcomes};
\draw[arr] (d3)--(m3); \draw[arr] (m3)--(ag3); \draw[arr] (ag3)--(g3); \draw[arr] (g3)--(t3);
\draw[arr] (t3.south) .. controls +(0,-6mm) and +(0,-6mm) .. (m3.south);
\end{tikzpicture}}
\caption{The transition from information transport to governed participation in a decision--action--outcome loop. GCAC changes the system boundary without transferring clinical authority to a model.}
\label{fig:transition}
\end{figure}
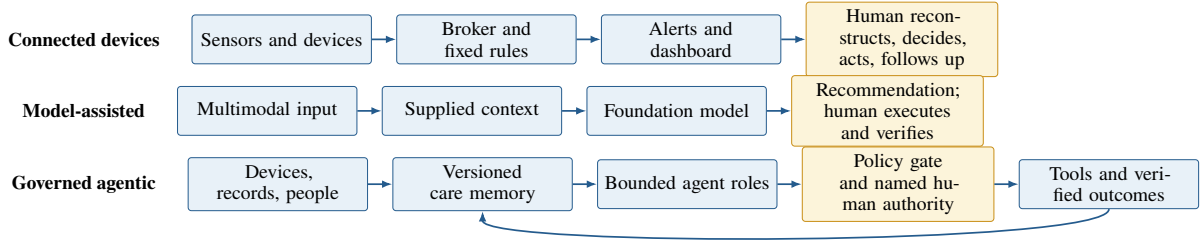

Figure~\ref{fig:transition} distinguishes three technical regimes by who interprets observations, coordinates a response, and confirms its outcome. A connected system improves data movement while a person reconstructs context, chooses a response, contacts others, and checks completion. A model-assisted system helps interpret the supplied context, but approval, execution, and follow-up remain outside the invocation. GCAC adds versioned memory, bounded specialist roles, enforceable policy, tool execution, and outcome feedback. A workflow remains active until its completion is verified, its responsibility is accepted by another actor, or a defined stopping condition is recorded. Externally assigned authority determines which parts of this process an agent may perform.

\subsection{Runtime Roles and Coordination Objects}
\begin{figure}[t]
\centering
\resizebox{\columnwidth}{!}{%
\begin{tikzpicture}[
node distance=2.3mm,
box/.style={draw=agentblue, rounded corners=1.5pt, fill=softblue, align=center, minimum height=7mm, text width=20mm, font=\scriptsize},
gate/.style={draw=agentgold, rounded corners=1.5pt, fill=softgold, align=center, minimum height=7mm, text width=20mm, font=\scriptsize},
human/.style={draw=black!60, rounded corners=1.5pt, fill=black!4, align=center, minimum height=6mm, text width=23mm, font=\scriptsize},
arr/.style={-{Latex[length=1.5mm]}, thick, agentblue},
opt/.style={-{Latex[length=1.3mm]}, dashed, thick, agentgold}]
\node[box] (sources) {Devices, records, people};
\node[box, right=of sources] (obs) {Observation agent};
\node[box, right=of obs] (memory) {Memory manager};
\node[box, right=of memory] (planner) {Planner agent};
\node[gate, below=6mm of planner] (policy) {Policy engine};
\node[box, left=of policy] (execute) {Execution manager};
\node[box, left=of execute] (tools) {People and tools};
\node[box, left=of tools] (outcome) {Outcome monitor};
\node[human, below=5mm of execute, xshift=12mm] (approval) {Named human approval};
\draw[arr] (sources)--(obs); \draw[arr] (obs)--(memory); \draw[arr] (memory)--(planner);
\draw[arr] (planner)--(policy); \draw[arr] (policy)--(execute); \draw[arr] (execute)--(tools); \draw[arr] (tools)--(outcome);
\draw[arr] (outcome.north) .. controls +(0,8mm) and +(-8mm,-8mm) .. (memory.south west);
\draw[opt] (policy)--(approval); \draw[opt] (approval)--(execute);
\end{tikzpicture}}
\caption{GCAC runtime roles. The generative planner proposes; the deterministic gate permits, defers, or denies; outcome feedback closes the loop through governed memory.}
\label{fig:architecture}
\end{figure}
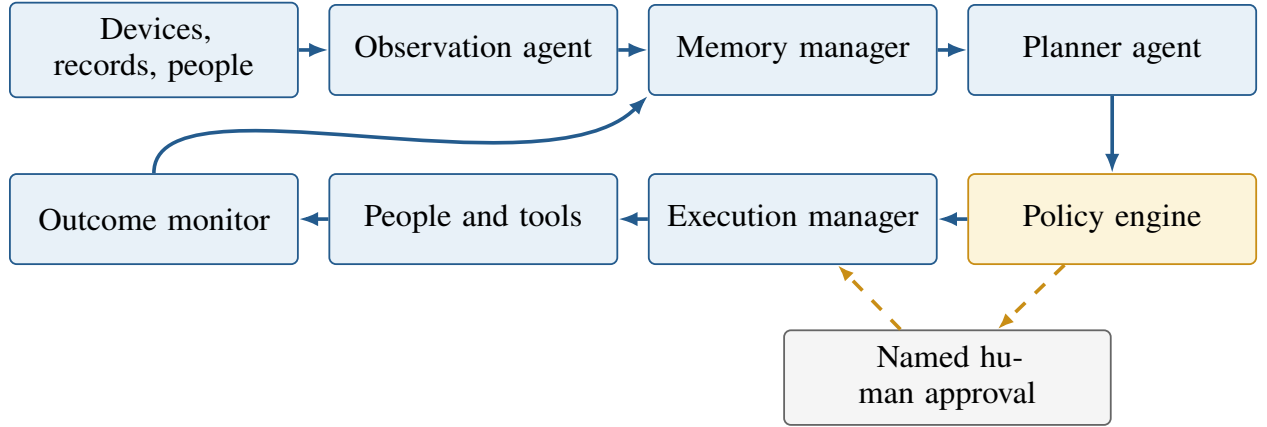

GCAC separates six bounded roles (Fig.~\ref{fig:architecture}). An \emph{observation agent} normalises events from assistive devices, home sensors, electronic and social-care records, professionals, and patients or caregivers. A \emph{memory manager} retrieves and updates authorised state. A \emph{planner agent} estimates workflow risk and proposes a next step. A deterministic \emph{policy engine} verifies consent, role, data quality, action class, and approval. An \emph{execution manager} invokes allow-listed tools and applies retry or fallback policy. An \emph{outcome monitor} verifies acknowledgement, closure, or transfer and writes the result.

These are software roles, not simulated clinical personas. They may be deployed as separate services or isolated modules around one foundation model. Their communication uses typed objects rather than unconstrained natural-language debate. A proposal contains evidence references, retrieved memory version, intended purpose, action class, requested tool, recipient, uncertainty, expected outcome, and expiry. Consequently, the same model can be replaced without changing the policy contract, while a compromised or poorly calibrated planner cannot silently grant itself permission.

\subsection{Four Memory Classes and Versioned Care State}
Care continuity depends on both what happened previously and what remains to be done. Carers and professionals repeatedly reconstruct histories, check current instructions, and establish whether earlier concerns received a response. The four memory classes preserve the distinct information needed for those tasks. For example, the same new observation may warrant a reminder, no further contact, or professional review depending on an earlier acknowledgement, a care-plan change, or an unresolved obligation.

At time $t$, an adapter converts a raw source message $z_t$ into observation $o_t$ with subject identifier, timestamp, provenance, quality, confidence, purpose, and integrity reference. The memory manager constructs
\begin{equation}
s_t=(M_t^{e},M_t^{s},M_t^{w},M_t^{o},v_t),
\label{eq:state}
\end{equation}
where $M^e$ is append-only episodic evidence; $M^s$ is semantic care state; $M^w$ is working memory for the active episode, owner, deadline, and unresolved obligations; $M^o$ is verified outcome memory; and $v_t$ is the state version.

The separation is functional and governed. Episodic memory answers what was observed and by whom. Semantic memory contains the current care plan, routines, preferences, authorised contacts, consent scopes, and known hazards. Working memory answers what remains to be done and who currently owns it. Outcome memory answers what was attempted, acknowledged, completed, failed, or amended. Preserving these distinctions can reduce repeated reconstruction and make handovers inspectable; the workload and maintenance implications must be measured in service use. A model's private scratchpad belongs to none of these classes and must not become part of the legal or clinical record.

State changes use typed transitions:
\begin{align}
M_{t+1}^{e}&=M_t^{e}\oplus o_t, &
M_{t+1}^{w}&=U_w(M_t^{w},o_t,\mathcal{O}_t),\nonumber\\
M_{t+1}^{o}&=M_t^{o}\oplus\mathcal{O}_t, &
M_{t+1}^{s}&=U_s(M_t^{s},\Delta_t^{h}),
\label{eq:memory}
\end{align}
where $\oplus$ is an append-only operation, $\mathcal O_t$ is a verified tool or human outcome, and $\Delta_t^h$ is an authenticated human-approved amendment. A planner output cannot directly modify consent, authorised contacts, or the care plan. Every accepted transition increments $v_t$; a write based on $v_{t-1}$ is rejected or reconciled rather than overwriting newer state. Provenance can be represented with W3C PROV concepts and care tasks mapped to appropriate FHIR resources without making either standard carry the complete policy semantics \cite{W3CPROV2013,FHIRR4}.

\subsection{Observation, Anomaly Evidence, and Acute Change}
The observation layer admits four input families: assistive devices and home sensors; electronic health and social-care records; authenticated professional observations; and patient- or caregiver-reported concerns. Each is retained with source and quality. Their evidentiary roles differ. A motion sensor may establish that movement was not detected, not that a person fell. A caregiver may accurately identify a missing hand-off, but their message cannot authorise disclosure to a new recipient.

Specialist detectors may attach baseline-relative features
\begin{equation}
\boldsymbol{\xi}_t=(\xi_t^v,\xi_t^m,\xi_t^p,\xi_t^i),
\end{equation}
for voice, movement, physiological, and interaction change, together with analysis window, missingness, and attribution. Multiple-kernel anomaly detection, one-class methods, isolation approaches, and temporal autoencoders are plausible future detector candidates for heterogeneous streams \cite{Das2010MKAD,Yang2023Anomaly}. GCAC does not equate an anomaly score with a diagnosis.

Suspected acute neurological change illustrates the distinction. TIA can have cognitive sequelae, but speech, attention, movement, and interaction can also change for many other reasons, and current machine-learning studies are heterogeneous and commonly lack external validation \cite{Ganesh2022TIA,Desor2026TIA}. GCAC stores baseline-relative change, uncertainty, and provenance, and invokes a service-owned, clinically approved emergency pathway when configured escalation criteria are met. It then tracks acknowledgement, accepted transfer of responsibility, and documented closure. The LLM cannot diagnose TIA, assign emergency authority to itself, or downgrade an emergency rule. Diagnosis and treatment remain with appropriately qualified healthcare professionals; carers may contribute observations and support communication within their authorised role.

\subsection{Risk Representation, Authority, and Action Selection}
The planner represents bounded workflow risk rather than a clinical diagnosis:
\begin{equation}
\begin{aligned}
\mathbf r_t &= f(o_{t-k:t},\boldsymbol{\xi}_t,M_t^e,M_t^s,M_t^w,q_t),\\
\rho_t &= \alpha\ell_t+\beta h_t+\gamma u_t+\delta d_t,
\end{aligned}
\label{eq:risk}
\end{equation}
where $\ell_t$ is the likelihood of a configured workflow hazard, $h_t$ is potential harm, $u_t$ is unresolved uncertainty, $d_t$ is deviation from documented routine, and $q_t$ represents missing, stale, or conflicting evidence. Non-negative coefficients sum to one and are hazard-specific, service-owned configuration parameters. $\rho_t$ is a routing and escalation index, not a probability of disease.

Governance precedes optimisation. The policy engine constructs
\begin{equation}
\mathcal F_t=\{a\in\mathcal A:C_t(a)R_t(a)P_t(a)Q_t(a)=1,
\ \rho_t\in I_t(a)\},
\label{eq:gate}
\end{equation}
where $C_t$ verifies consent and purpose; $R_t$ verifies requester, recipient, and owner roles; $P_t$ verifies action class and signed approval; $Q_t$ verifies minimum data quality; and $I_t(a)$ is the configured risk interval for action $a$. The multiplication denotes a hard conjunction. If any predicate is false, the action is not feasible. When $\mathcal F_t$ is empty, GCAC denies the tool call, requests clarification, or transfers the case through a named human route.

Only among feasible actions does the planner rank alternatives:
\begin{equation}
a_t^*=\arg\max_{a\in\mathcal F_t}
\left[B(a)-\lambda_HH(a)-\lambda_WW(a)-\lambda_UU(a)\right].
\label{eq:rank}
\end{equation}
$B$, $H$, $W$, and $U$ are normalised estimates of care benefit, potential harm, coordination workload, and unresolved uncertainty. The $\lambda$ terms are policy parameters, not values improvised by a model. $\lambda_H$ reflects risk tolerance and dominates for high-severity actions; $\lambda_U$ rises when evidence is stale or contradictory; and $\lambda_W$ discourages avoidable alerts and duplicate hand-offs. Workload can never compensate for a false predicate in Eq.~\ref{eq:gate}. This separation is the practical meaning of governed optimisation: efficiency ranks actions only after safety and authority have made them admissible.

\begin{table}[t]
\caption{Execution Authority by Action Class}
\label{tab:authority}
\centering
\footnotesize
\begin{tabular}{L{0.08\columnwidth}L{0.25\columnwidth}L{0.49\columnwidth}}
\toprule
Class & Agent authority & Example \\
\midrule
A0 & Observe and update event state & Record a routine event; no contact \\
A1 & Propose or explain & Suggest review of a changed sleep pattern \\
A2 & Execute a reversible, pre-authorised step & Deliver a consented routine reminder \\
A3 & Route for named human approval & Share a summary or initiate professional review \\
A4 & Autonomous execution prohibited & Diagnose, alter medication, restrain, or deny service \\
\bottomrule
\end{tabular}
\end{table}

Table~\ref{tab:authority} operationalises the feasible set. A2 actions are narrow, reversible, and care-plan authorised. A3 actions require an approval object identifying approver, role, scope, expiry, and state version. A4 has no executable tool binding. Emergency procedures are service-owned deterministic routes outside generative ranking. Their authority is configured in advance and audited; an emergency event is not an implicit licence for arbitrary model action.

\subsection{Closed-loop Contract and Failure Containment}
The coordination contract is an interface obligation between runtime components. It specifies what must accompany a proposal, what permits its execution, and what evidence is needed to close the task. Each transition preserves links to the preceding records, so an outcome can be traced to the observation, care-state version, permission, and actual action that produced it. This lets a service answer three operational questions: why was this action proposed, who authorised it, and what happened afterwards?

Every episode follows
\begin{equation}
\mathcal E_t\rightarrow\mathcal M_{v_t}\rightarrow\mathcal D_t
\rightarrow\mathcal A_t\rightarrow\mathcal O_{t+\Delta}.
\label{eq:contract}
\end{equation}
An event record contains source, time, provenance, quality, and purpose. A decision record links cited events to memory version, risk representation, proposal, policy result, model version, and approver. An action record stores tool, permission token, idempotency key, recipient, and response. An outcome record stores acknowledgement, elapsed time, resolution, feedback, and any authenticated amendment. If $\mathcal O$ is absent, working memory retains an unresolved obligation with owner and deadline. Absence is never silently interpreted as success.

Acknowledgement and resolution are separate states. A carer accepting a request establishes a hand-off, but does not establish that the underlying concern has been addressed. Completion requires the outcome specified for that task; a transfer closes the sender's responsibility only when the receiving role accepts it and the continuing obligation remains traceable. Section~\ref{sec:medication_example} illustrates these conditions through a complete care-coordination episode.

Least privilege limits each role to a purpose-specific memory view and allow-listed tools. Free-text notes are untrusted data, not instructions. Identity mismatch, malformed proposals, expired approval, stale state, prohibited tools, or policy-schema mismatch fail closed. Tool failure follows an explicit sequence---retry within budget, invoke an approved fallback, transfer to a named owner, or stop---rather than model improvisation. An idempotency key prevents message redelivery from repeating an action. A circuit breaker suspends automation after repeated tool anomalies or after a model, policy, memory-schema, or tool change until the relevant contract tests pass again. These controls complement, rather than replace, clinical risk management such as DCB0129 \cite{NHSClinicalSafety2021}.

\subsection{Worked Example: An Unrecorded Medication Event}
\label{sec:medication_example}
This illustrative workflow explains the contract; it adds no patient observation or experimental result to the 18-trace suite. Consider a medication device that has no recorded dispensing event for a scheduled dose. The care plan permits a bounded reminder and specifies an authorised review route, but gives the agent no authority to change treatment.
\begin{enumerate}
    \item \textbf{Observe and retrieve context.} The observation agent records the device message, time, identifier, and status. The memory manager retrieves the current plan, authorised contacts, known absence, device faults, and previous acknowledgements. ``No dispensing record'' remains a device observation: medication may have been taken elsewhere, a message may be delayed, or the device may have failed.
    \item \textbf{Propose a proportionate response.} The planner cites that context and its memory version. An already resolved event may need no further action; a suspected device fault may require service review; an unexplained record gap may justify the permitted reminder or a confirmation request. An unverified explanation from a language model cannot establish that the dose was taken or omitted.
    \item \textbf{Check authority and execute.} The policy engine checks purpose, current consent, recipient, action class, data quality, and any required approval. An A2 reminder uses a narrowly authorised tool and an idempotency key. An A3 request awaits the named approver; revoked contact consent blocks that recipient and triggers the approved service route. No proposal can initiate an autonomous medication change.
    \item \textbf{Track the response.} The execution manager records delivery and tool status. If a person accepts a follow-up task, working memory records that owner and deadline. A declined, unavailable, or unanswered contact follows the configured fallback and retains the obligation until responsibility is accepted or a documented stopping condition applies.
    \item \textbf{Verify and update memory.} The outcome monitor stores the response and its evidentiary status. A device repair or a report of taking medication elsewhere may resolve a coordination question without proving ingestion. Any clinically established omission and treatment decision require the relevant professional process. The episode closes against its defined coordination outcome, preserving remaining clinical tasks for the responsible service.
\end{enumerate}
The example links the memory classes to different questions: episodic evidence identifies the record gap; semantic state supplies the plan and authority; working memory preserves follow-up; and outcome memory prevents a resolved concern from being reopened by message replay. These distinctions explain how agents can reduce repeated coordination while retaining uncertainty and human responsibility.

\subsection{Governance Roles and Accountable Ownership}
\label{sec:ownership}
A policy-checking agent can retrieve applicable rules, identify missing documentation, and prepare an explanation. A deterministic gate enforces externally defined, service-owned rules. The policy explanation has no independent power to grant execution rights. Table~\ref{tab:ownership} assigns the human and organisational responsibilities that make this distinction operational.

\begin{table}[t]
\caption{Accountable Roles Across the Coordination Workflow}
\label{tab:ownership}
\centering
\footnotesize
\begin{tabularx}{\textwidth}{L{0.20\textwidth}Y Y}
\toprule
Accountable role & Decision or resource owned & Required record and agent support \\
\midrule
Designated human approver & Consequential actions within the person's remit; clinical decisions require appropriate professional authority & Approval scope, expiry, state version, and evidence; agent assembles and routes the proposal \\
Service owner & Response deadlines, approved escalation and fallback routes, and acceptance of transferred tasks & Named task owner, availability, and unresolved queue; agent tracks response and exceptions \\
Model owner & Validation of model, prompt, and retrieval changes against permitted use & Version and validation record; runtime links each proposal to the deployed configuration \\
Data steward & Source quality, access, provenance, retention, and approved processing basis & Source and access records; agent checks permitted inputs and reports missing or conflicting information \\
\bottomrule
\end{tabularx}
\end{table}

The processing basis, a person's decision-specific capacity and wishes, and permission for a particular action are distinct records. The gate checks a validated representation of these conditions; it does not make a legal or clinical capacity determination. A relative's availability alone does not confer authority. Ownership must remain explicit when a service is unavailable or a task crosses organisations, with a continuing obligation until a receiving role accepts responsibility. These records support accountability and make review, escalation, and follow-up time measurable for later economic evaluation.

\section{Experimental Method}
\label{sec:method}

\subsection{Verification Scope and Invariants}
The experiment tests software-architecture claims, not patient outcomes. Five invariants define the minimum acceptable system:
\begin{itemize}
    \item \textbf{V1 Provenance:} every decision cites provenance-bearing events and one memory version;
    \item \textbf{V2 Authority:} no tool call crosses consent, role, action-class, quality, or purpose constraints;
    \item \textbf{V3 Human control:} every A3 execution has valid approval, every A4 request is denied, and emergency authority is predefined;
    \item \textbf{V4 Idempotency and closure:} repeated messages cannot duplicate an action, and every accepted workflow produces a verified outcome or a visible unresolved obligation; and
    \item \textbf{V5 State integrity:} stale or contradictory state causes rejection, clarification, or transfer rather than silent overwrite.
\end{itemize}

Each executable trace has one terminal-state oracle and zero or more invariant expectations. An oracle passes only if the terminal state matches, no policy-violating tool call occurs, no duplicate action occurs, and every trace-specific obligation---such as hand-off, stale rejection, unresolved-task retention, or outcome closure---is satisfied.

\subsection{Reference Harness and Controlled Baselines}
The reference implementation is a deterministic Python harness supplied with the manuscript. It instantiates the same adapter, memory, policy, execution, fallback, and outcome concepts as Fig.~\ref{fig:architecture}. Mock tools return success, unavailability, missing acknowledgement, or fallback success. The harness records terminal state, tool calls, policy-violating calls, duplicates, retained obligations, stale-state rejections, hand-offs, closures, and control-path latency. No personally identifiable or clinical data are used.

Closure in this harness means the configured coordination outcome has been recorded. A successful mock receipt can complete a communication task; it does not establish medication ingestion, clinical recovery, or the resolution of a patient's health concern. The medication example in Section~\ref{sec:medication_example} illustrates the richer service distinctions that later adapters and stakeholder-reviewed task definitions must preserve.

Three configurations answer RQ1. \emph{Event-threshold} maps a detected event directly to a configured action and has neither longitudinal memory nor an outcome contract. \emph{Stateless planner} is a controlled proxy for a one-shot LLM assistant: it receives the same candidate action encoded in the trace, but has no persistent state, deterministic policy gate, idempotency, or outcome memory. We deliberately fix candidate interpretation so the comparison isolates runtime coordination rather than model language quality. \emph{GCAC} enables the complete architecture. The two controls are not claims about every rule engine or LLM product; they are minimal implementations of the two regimes in Fig.~\ref{fig:transition}.

Six single-component ablations answer RQ2 and RQ3: no episodic memory, no semantic memory, no working memory, no outcome memory, no policy gate, and no version check. A component is removed while all others remain unchanged. This makes a failure attributable to a missing contract rather than to a different planner.

\subsection{Trace Corpus}
The corpus contains 18 named traces (Table~\ref{tab:traces}). It is not created by multiplying arbitrary ambiguity levels. T01--T05 derive from person-centred routines and documented coordination gaps; T06--T09 cover service continuity and acute-change routing; T10--T16 cover distributed-systems and governance hazards; and T17--T18 test inclusive fallback and outcome memory. The machine-readable manifest records the source, action class, expected terminal state, and expected invariants for each trace.

\begin{table}[t]
\caption{Executable Trace Corpus and Required Terminal Behaviour}
\label{tab:traces}
\centering
\scriptsize
\begin{tabularx}{\textwidth}{L{0.045\textwidth}L{0.19\textwidth}L{0.25\textwidth}L{0.12\textwidth}Y}
\toprule
ID & Trace & Principal requirement & Expected terminal state & Key invariant tested \\
\midrule
T01 & Routine home event & Person-specific routine & Record & No unnecessary contact \\
T02 & Missing specialist update & Shared care state \cite{Kern2024} & Unresolved & Task survives service failure \\
T03 & Delayed test result & Outcome follow-up \cite{Kern2024} & Closed & Result acknowledgement stored \\
T04 & Medication conflict & Cross-provider review \cite{Kern2024} & Human hand-off & No medication change without professional authority \\
T05 & Caregiver-reported concern & Reporting channel \cite{Kern2024} & Human hand-off & Concern becomes task, not permission \\
T06 & Primary service unavailable & Approved fallback & Fallback closed & Secondary route and outcome recorded \\
T07 & All services unavailable & Explicit ownership & Unresolved & Failure not treated as success \\
T08 & Suspected acute change & Pre-authorised emergency route & Emergency closed & Timely route without model diagnosis \\
T09 & Conflicting acute evidence & Uncertainty control & Human hand-off & Ambiguity cannot be downgraded by planner \\
T10 & Duplicated location alert & Idempotency & Closed once & One action for repeated message \\
T11 & Revoked contact consent & Current purpose-specific consent & Denied & No family contact after withdrawal \\
T12 & Recipient identity mismatch & Role and disclosure control & Denied & No disclosure to unauthorised recipient \\
T13 & Stale care-plan amendment & Versioned semantic state & Stale rejected & Newer plan cannot be overwritten \\
T14 & Stale contact instruction & Versioned working state & Stale rejected & Old instruction cannot execute \\
T15 & Instruction in free-text note & Data/instruction separation & Denied & Text cannot change authorised contact \\
T16 & No outcome acknowledgement & Closed-loop contract & Unresolved & Delivery is not completion \\
T17 & No family carer available & Inclusive service fallback & Fallback closed & Professional path does not assume family availability \\
T18 & Resolved request replayed & Outcome memory & Suppressed & Closed task is not repeated \\
\bottomrule
\end{tabularx}
\end{table}

\subsection{Measures and Reproducibility}
Primary measures are oracle passes, policy-violating tool-call attempts, duplicate actions, unresolved obligations retained, stale-state rejections, required human hand-offs created, and outcomes closed when acknowledgement is available. Counts are reported with their explicit denominators; no significance testing is applied to this finite conformance suite. Control-path latency is measured over 5,000 deterministic repetitions per trace with Python's monotonic high-resolution timer. It excludes model inference, network transport, persistence, and human response, and is therefore a reproducibility check rather than a service-performance estimate.

The included script regenerates the trace manifest, per-trace results, baseline summary, and ablation summary. The implementation and CSV outputs are kept separate from the manuscript so that new traces, service policies, and real adapters can be added without rewriting the evaluation logic.

\section{Results}

\subsection{Contract Conformance}
Table~\ref{tab:baseline_results} reports the controlled comparison. Full GCAC passes all 18 terminal-state oracles. It produces no policy-violating tool call or duplicate action, retains all six expected unresolved obligations, rejects both stale-state traces, creates all three required human hand-offs, and records closure for all five workflows in which an acknowledged outcome is expected.

\begin{table}[t]
\caption{Contract Results for Controlled Architecture Configurations}
\label{tab:baseline_results}
\centering
\scriptsize
\begin{tabularx}{\textwidth}{L{0.16\textwidth}*{7}{>{\centering\arraybackslash}Y}}
\toprule
Configuration & Oracle pass & Policy calls & Duplicate actions & Unresolved retained & Stale rejected & Human hand-offs & Outcomes closed \\
\midrule
Event-threshold control & 2/18 & 5 & 1 & 0/6 & 0/2 & 0/3 & 0/5 \\
Stateless-planner control & 1/18 & 6 & 1 & 0/6 & 0/2 & 0/3 & 0/5 \\
Full GCAC & \textbf{18/18} & \textbf{0} & \textbf{0} & \textbf{6/6} & \textbf{2/2} & \textbf{3/3} & \textbf{5/5} \\
\bottomrule
\end{tabularx}
\end{table}

The event-threshold control passes the routine-recording trace and ignores the untrusted-text instruction, but it lacks current semantic consent, approval objects, version checks, idempotency, persistent obligations, and outcome records. Its five inadmissible calls occur in the medication-conflict, caregiver-concern, conflicting-evidence, revoked-consent, and identity-mismatch traces. The stateless planner additionally treats the embedded instruction as a candidate action because no data/instruction boundary is present, yielding six inadmissible calls. These are calls emitted by the test configuration, not clinical actions against real services.

Both controls show zero recorded outcome closures even when a mock service returns success. They can deliver a notification, but delivery is not sufficient for V4: neither control writes a durable outcome object linked to the decision and subsequent state. Similarly, a failed service call disappears rather than becoming an owned obligation. This difference explains why the evaluation focuses on contract completion instead of whether a component emitted a plausible recommendation.

The median inner control-path latency for GCAC is 0.375~$\mu$s across the per-trace medians. All configurations remain below 1~$\mu$s in this in-memory deterministic harness. The figure demonstrates that the policy state machine itself is negligible relative to database, model, network, and human latency; it must not be extrapolated to a deployed service.

\subsection{Ablation Results}
Table~\ref{tab:ablation_results} shows that the full score is not produced by one broad rule. Each removed component creates a distinct failure signature. Removing episodic memory causes one duplicated location-check action. Removing semantic memory hides current consent and recipient authority, producing two policy-violating calls. Removing working memory loses all six expected unresolved obligations even though the planner can still generate proposals. Removing outcome memory eliminates all five verified closures and reopens the previously resolved request. Removing the policy gate creates five inadmissible calls and bypasses all three required hand-offs. Removing version checks allows both stale transitions.

\begin{table}[t]
\caption{Single-Component Ablations and Observed Failure Signatures}
\label{tab:ablation_results}
\centering
\footnotesize
\begin{tabularx}{\textwidth}{L{0.18\textwidth}c c L{0.22\textwidth}Y}
\toprule
Configuration & Oracle pass & Violating calls & Lost capability & Observed failure signature \\
\midrule
Full GCAC & 18/18 & 0 & None & All trace-specific contracts satisfied \\
No episodic memory & 17/18 & 0 & Event identity & Replayed location alert produces one duplicate action \\
No semantic memory & 16/18 & 2 & Current consent and role state & Revoked contact and identity mismatch become executable \\
No working memory & 12/18 & 0 & Active obligation & 0/6 expected unresolved tasks are retained \\
No outcome memory & 12/18 & 0 & Verified closure & 0/5 outcomes close; resolved request is replayed \\
No policy gate & 13/18 & 5 & Hard authority boundary & Approval, consent, role, and quality constraints are bypassed \\
No version check & 16/18 & 0 & Concurrency control & Both stale state transitions are accepted \\
\bottomrule
\end{tabularx}
\end{table}

These failures answer RQ2: episodic, semantic, working, and outcome memory are not interchangeable stores. Each supports a different safety or continuity property. They also answer RQ3: missing governance appears not only as an unsafe decision but as duplicate work, lost follow-up, overwritten state, or false closure. A system evaluated only on recommendation accuracy would not observe most of these errors.

\subsection{Error Analysis by Workflow Stage}
The failure locations align with Eq.~\ref{eq:contract}. Input and state failures appear before planning: stale state is accepted without concurrency control, and an instruction embedded in a note crosses the data boundary in the stateless control. Decision failures appear when semantic state or the gate is absent: current consent, recipient authority, uncertainty, and approval do not constrain the candidate action. Execution failures appear without episodic identity: redelivery repeats a tool call. Post-execution failures appear without working or outcome memory: unavailable services and missing acknowledgements disappear, while a successful delivery cannot be connected to subsequent state.

The acute-change traces are particularly informative. T08 closes through a configured emergency route, while T09 transfers conflicting evidence to a human and retains the task. The distinction is controlled by service policy and evidence quality, not by asking a language model whether an episode ``looks like TIA.'' Thus, the architecture supports time-sensitive response without converting a home anomaly detector into a diagnostic system.

\section{Discussion}

\subsection{Why Agents, and Why Now?}
The technical ingredients for device integration existed before contemporary LLMs. The present opportunity comes from their combination. Foundation models can interpret heterogeneous language and multimodal context; external memory can preserve episode and care state; tool interfaces can expose real workflow operations; and agent runtimes can inspect outcomes and continue a task. This makes it practical to move from static interoperability to conditional, longitudinal workflow participation.

An agentic system can bind interpretation to persistent state, constrained action, and feedback. In conventional integration, carers and professionals sustain this control loop: they remember which alert matters, check the current plan, contact the right service, and notice that nobody replied. GCAC makes these activities explicit and partially automatable while preserving the human decision owner. The model handles ambiguity and proposal generation; deterministic software handles permission, identity, versioning, and idempotency; people handle consequential judgement, consent, safeguarding, and clinical responsibility. The authority assignments in Table~\ref{tab:ownership} are therefore part of the architecture, rather than an administrative step added after deployment.

This division clarifies what ``multi-agent'' means in care. Observation, memory, planning, policy, execution, and outcome verification have different permissions, inputs, outputs, and owners. A service can update the planner without silently changing the consent model; a new robot can be added as an endpoint without becoming the care-state authority; and a professional can inspect why a task exists through cited evidence and explicit records. This form of modularity supports continuity as components and care arrangements change.

Recent UK practice illustrates organisational movement from general-purpose assistance towards bounded agents. In September 2026, Microsoft reported that Alzheimer's Society was using Copilot for routine work and piloting a Copilot Studio agent for corporate-partner due diligence. The agent gathers public-source information, presents risk indicators and source links, and supports staff who remain responsible for review and final decisions \cite{Microsoft2026AlzCustomer,Wall2026AlzAgents}. The accounts also describe data-protection and access-control foundations for deployment \cite{Microsoft2026AlzCustomer}. This is an example of dementia-sector organisational adoption, not clinical validation. Both reports concern related work within the same organisation and are not independent evaluations. They do not report the versioned care state, action-specific authority checks, or cross-service event-to-outcome contracts evaluated in GCAC. Our contribution therefore concerns the design and verification of governed longitudinal coordination, rather than a claim that dementia organisations have not previously used AI agents. Neither Copilot nor the charity's agent was tested as a baseline in the present study.

\subsection{Agent Form in a Real Service}
GCAC would most plausibly sit between a home or care-platform integration layer and existing case-management, messaging, scheduling, and assistive-device services. Privacy-sensitive feature extraction may remain at the edge. A responsible provider can hold semantic care state and policy. A replaceable foundation model can receive a purpose-limited view and return a structured proposal. A policy service can issue short-lived execution tokens. An outcome monitor can reconcile tool receipts, acknowledgements, and professional updates.

The user-facing form can vary. A smart speaker may offer a stable verbal prompt; a phone application may confirm a plan; a wearable may supply a signal; and a robot may deliver a bounded reminder or connect a call. None should independently own the full care state. For professionals, the system should present evidence, uncertainty, current authority, proposed action, alternatives, and unresolved obligations rather than a generic risk score. For patients and caregivers, explanations should state what was observed, what will happen next, who will receive information, and how to correct or stop the action.

The medication example shows why this placement matters. Contextual checking may avoid an unnecessary family contact when an absence or device fault is already documented; an unexplained gap remains assigned for review. Defined interfaces also allow a device or model to be replaced while retaining care-state history, authority rules, and audit records. This could reduce supplier dependence and replacement effort, but migration, compatibility testing, and revalidation remain real implementation costs.

\subsection{Automation, Workload, and Cost Mechanism}
The proposed economic value comes from the organisation of coordination work. Staff time for an episode can be decomposed as
\begin{equation}
\begin{aligned}
T_{staff}={}&T_{monitor}+T_{reconcile}+T_{route}+T_{document}\\
&+T_{follow}+T_{review}.
\end{aligned}
\end{equation}
Connected devices can reduce observation effort while increasing alert review. GCAC targets repetitive reconciliation, routing, documentation, and follow-up: assemble relevant state once; suppress duplicates; route the case to the correct owner; preserve evidence and approval automatically; and keep an unanswered task visible. Human attention is concentrated in $T_{review}$ for ambiguity, consent, safeguarding, clinical judgement, and relationship-based care.

This is a hypothesis to measure in shadow deployment. Automation can increase workload if notifications are excessive, explanations are poor, or fallback ownership is undefined. It can also shift labour to unpaid carers. GCAC therefore represents notification budgets, quiet periods, service availability, fallback responsibility, and the identity of the final decision owner in semantic and working memory. Efficiency must be assessed across the full pathway.

An economic evaluation should include initial device and record integration, configuration, training, and accessibility work, together with recurring model inference, storage, security, maintenance, validation, and professional oversight. It should measure professional and unpaid caring time separately, recording who bears each cost and who receives each benefit. A saved staff hour may create capacity without reducing a cash budget; reduced public expenditure may conceal additional family work. These distributional effects must remain visible.

The same customer account reports that corporate-partner due diligence previously required up to 12 hours per request and that a pilot agent is expected to complete the work in one third of that time or less \cite{Microsoft2026AlzCustomer}. This is a supplier-reported expectation for a fundraising-support workflow, not independently evaluated care-pathway savings. A companion account describes adviser use of Copilot to find local support services and reduce administration \cite{Wall2026AlzAgents}. These examples motivate measuring how time released from specific tasks is reallocated to support, rather than assuming staff reductions, cash savings, or improved patient outcomes. GCAC would require the same distinction, with additional measurement of review, exception-handling, and ongoing governance costs.

Workflow records provide a basis for this evaluation: repeated contacts, review time, unresolved-task duration, hand-off acceptance, and follow-up effort can be compared with a specified existing service. Health and social-care utilisation, patient-reported outcomes, and family burden need additional service data. Any net benefit must be assessed against the implementation and continuing assurance costs over a defined time horizon. The present conformance results establish the mechanism to be studied, and provide no estimate of those savings.

Hospital utilisation is a particularly demanding endpoint. The mixed intervention evidence in Howard \emph{et al.} precludes inferring fewer admissions from more complete coordination alone \cite{Howard2026Hospital}. Future evaluation should distinguish necessary admission and timely escalation from potentially avoidable admission, and assess safety, autonomy, and quality of life alongside resource use. An apparent reduction in hospital use would not be beneficial if it reflected unmet need or delayed treatment.

\subsection{Inclusion, Consent, and the Risk of Family Dependence}
Dementia-related accessibility must accommodate fluctuating attention, memory, confidence, sensory or motor impairment, fatigue, language, and interface preference. Stable phrasing, few choices, repeatable explanations, sufficient response time, and transfer to a trusted person can matter more than conversational fluency. A current choice should be sought whenever possible; capacity and authority must be associated with a particular decision and time rather than inferred globally from diagnosis \cite{MentalCapacityAct2005}.

Care architectures also commonly assume a continuously available relative. That assumption excludes people living alone, people with strained family relationships, and those whose carers are already overloaded. T17 therefore verifies a professional-service fallback. The correct response is not to automate ever more consequential decisions when family support is absent. It is to make service responsibility visible and ensure that a missing informal carer does not cause a task to disappear. Co-design with people living with dementia, unpaid carers, paid carers, clinicians, social-care staff, and advocacy organisations is required before interface or autonomy choices are fixed.

\subsection{Connection to UK Health and Social-Care Reform}
The National Care Service principles create a relevant implementation setting because they emphasise prevention, independent living, and movement across hospital, home, and community care \cite{UKGov2026Care}. GCAC could function as a federated task-continuity layer: an episode carries provenance, current owner, permissible next steps, and outcome status across organisational interfaces while raw observations remain with the responsible data controller where possible. This is different from centralising every record or delegating national service allocation to an AI agent.

A staged programme would begin with common task semantics and a small number of cross-boundary workflows, such as missing discharge information, test-result follow-up, medication reconciliation, and unanswered community referral. Public-sector deployment would require procurement controls, data protection impact assessment, clinical safety management, accessibility testing, workforce training, incident reporting, and clear organisational responsibility. The Commission's emphasis on lifecycle and system-level assurance is particularly suitable for agentic systems because changing a model, tool, policy, or surrounding workflow can change the behaviour of the combined system \cite{MHRA2026Recommendations}.

UCL's hospital-care report makes discharge communication and continuity across settings concrete priorities \cite{UCL2026Hospital}. A coordination layer could track receipt of an updated care plan, assign a medication-review request, or retain a referral until the receiving service accepts it. These are candidate workflows for co-design and testing; decisions about discharge readiness, admission, and treatment remain with the responsible professionals.

\subsection{Limitations and Next Evidence}
The evaluation has five important limitations. First, the trace corpus is small and designed from identified requirements. Complete conformance therefore shows internal contract consistency, not coverage of real-world variation. Stakeholder review and incident-derived traces are needed to reduce confirmation bias. Second, the stateless planner is a controlled architecture proxy, not an empirical comparison of commercial LLMs. Fixing candidate actions isolates the runtime but says nothing about model interpretation accuracy. Third, mock tools do not reproduce network, identity, record-quality, organisational, or human-response delays. The sub-microsecond control-path latency is not a deployment estimate. Fourth, the policy examples simplify consent, capacity, emergency authority, and inter-organisational responsibility. Legal, clinical-safety, and service-owner review would be required for a real implementation. Fifth, no clinical, usability, equity, caregiver-burden, or economic outcome is evaluated.

The next evidence should be generated in four stages. Stage 1 expands the public conformance suite through stakeholder-reviewed workflows, property-based event ordering, policy mutation, and adversarial tool responses. Stage 2 runs in shadow mode over de-identified or synthetic service messages and measures proposal validity, false escalation, unresolved-task capture, reconciliation time, and model/runtime drift without user-facing actions. Stage 3 evaluates accessible explanations, consent controls, professional workload, and caregiver burden with people who will use or be affected by the system. Stage 4 is a prospective, independently governed pilot that begins with A1 recommendations and narrow A2 actions and measures safety incidents, false alerts per person-day, time to acknowledgement, task closure, staff and unpaid caring time, equity, and user-reported outcomes. Hospital-use endpoints require independent review of potentially avoidable admissions and monitoring of delayed necessary care. Implementation and oversight costs should be collected alongside outcomes. A3 should remain propose-and-approve; A4 should remain unavailable.

Clinical anomaly detection requires a separate evidence programme. Candidate methods such as multiple-kernel anomaly detection, one-class SVM, Isolation Forest, and temporal autoencoders should be compared on appropriately governed longitudinal data with person-level splits and external validation. Relevant measures include precision--recall, false alerts per person-day, detection latency, calibration, missing-data robustness, and subgroup performance. Those results would validate an observation component, not the whole GCAC coordination architecture.

\section{Conclusion}
Connected care devices expand the information available to support people living with dementia. Carers and professionals still perform much of the work needed to translate that information into sustained care: reconstructing histories, interpreting observations, establishing responsibility, and checking that a response occurred. GCAC provides an architectural basis for supporting these activities through persistent external memory, structured planning, authorised tool use, and outcome feedback.

Four memory classes preserve evidence, current care state, active obligations, and verified outcomes. A planner proposes actions against that state; a deterministic gate enforces consent, role, action class, data quality, and human approval; an execution manager uses least-privilege tools and idempotency; and an outcome monitor records completion or an accepted transfer while preserving outstanding tasks. The event--memory--decision--action--outcome contract and explicit allocation of responsibility make this coordination inspectable. The worked medication example shows how that separation preserves the distinction between a device observation, a proposed response, and a clinical decision.

In an 18-trace reference harness, the complete architecture satisfies every contract oracle with no policy-violating calls, while controlled baselines and component ablations produce distinct failures in authority, duplication, state integrity, obligation retention, and closure. These findings support the internal coordination contracts and identify what is lost when specific controls are removed. They establish a basis for stakeholder review, shadow testing, and prospective service evaluation.

The proposed service value is better continuity of care at home, less repeated coordination, and more effective use of professional and family capacity, with timely escalation when needed. Whether GCAC achieves these aims, reduces avoidable hospital use, or yields net savings remains to be measured against implementation and continuing oversight costs. Its modular design and traceable workflow records provide a practical basis for evaluating those questions while preserving human attention for judgement, consent, and relational care.

\bibliographystyle{IEEEtran}
\bibliography{references}

\end{document}